\documentclass[a4paper,12pt]{iopart}
\usepackage[cp1250]{inputenc}
\usepackage[]{amssymb}
\usepackage[]{amsthm}
\usepackage[]{mathrsfs}
\usepackage[]{tensor}
\usepackage[]{eucal}
\usepackage{bm}
\usepackage{upgreek}
\usepackage[symbol]{footmisc}
\renewcommand{\d}{\mathrm{d}}

\newcommand{\bet}{\bm{\beta}}
\newcommand{\sF}{{*F}}
\newcommand{\sJ}{{*J}}
\newcommand{\qqd}{\ , \quad}
\newcommand{\txand}{\quad \textrm{and} \quad}

\newcommand{\be}{\begin{equation}}
\newcommand{\ee}{\end{equation}}
\newcommand{\bea}{\begin{eqnarray}}
\newcommand{\eea}{\end{eqnarray}}
\newcommand{\0}{\nonumber}

\newcommand{\fv}[1]{\mathnormal{#1}}

\begin{document}

\begin{flushright}
TZF-13-01
\end{flushright}

\title[Subtleties of Invariance]{Subtleties of invariance, covariance and observer independence}

\bigskip

\author{B Klajn$^a$ and I Smoli\'c$^b$}

\address{$^a$ Theoretical Physics Division, Rudjer Bo\v skovi\'c Institute, P.O.Box 180, HR-10002 Zagreb, Croatia}
\address{$^b$ Department of Physics, Faculty of Science, University of Zagreb, p.p.~331, HR-10002 Zagreb, Croatia}

\eads{\mailto{bruno.klajn@irb.hr}, \mailto{ismolic@phy.hr}}

\vspace{20pt}

\begin{abstract}
The role of the observers is frequently obscured in the literature, either by writing equations in a coordinate system implicitly pertaining to some specific observer or by entangling the invariance and the observer dependence of physical quantities. Using examples in relativistic kinematics and classical electrodynamics, we clarify the confusion underlying these misconceptions.
\end{abstract}

\pacs{03.30.+p, 03.50.De}


\vspace{20pt}

\section{Introduction}

The clarity of exposition of conceptual issues in physics is always endangered by words whose precise meaning is obscured by their overuse. Two known examples are the adjectives \emph{invariant} and \emph{covariant}. To say that some physical quantity $\mathcal{A}$ is invariant means that its value does not change with respect to some (coordinate or gauge) transformation. For example, any physical observable represented by a scalar quantity, such as temperature or pressure, is invariant with respect to coordinate transformations. To say that some physical quantity $\mathcal{A}$ transforms \emph{covariantly} means that it transforms in the same way as some other physical quantity $\mathcal{B}$. Usually, this refers to physical quantities represented by tensors and appearing in tensorial equations, such as Maxwell's electrodynamic or Einstein's gravitational field equations. A covariant transformation property guarantees that both sides of the equation will retain their form in every coordinate system, the only technical change being the appearance of the primes on all the indices, indicating the change of the coordinate system.

\bigskip

However, qualifying some quantity as \emph{noninvariant} or \emph{noncovariant} could be seen as a ``red herring''. Namely,  many physical observables can be written in a manifestly invariant or tensorial form, at the expense of introducing the observer's 4-velocity $\fv{o}^a$, as will be thoroughly illustrated in the examples below.\footnote{Here, 4-velocity plays the role of the basis time-like 4-vector of a given reference frame.} This may seem to be mere nitpicking, but by ignoring this subtlety one can be drawn into erroneous reasoning. In this paper we shall advocate the necessity of maintaining the fine distinction between invariance and observer independence in order to avoid possible conceptual pitfalls.

\bigskip

Before proceeding with more concrete examples, we shall make several remarks about the notation. We work in SI system of units so that all ``$c$'' factors (the speed of light) are always kept explicit in the expressions. Furthermore, in order to keep track of different conventions, we shall keep the choice of the metric signature of the Minkowski spacetime explicit,
\be
\eta_{\mu \nu} = \eta\,\mathrm{diag}\left( 1,-1,-1,-1 \right) \qqd \eta \equiv \eta_{00} = \pm 1 \, .
\ee
This means that the square of the 4-velocity $\fv{v}^a$ and 4-momentum $\fv{p}^a$ is given by
\be
\fv{v}^a \fv{v}_a = \eta c^2 \qqd \fv{p}^a \fv{p}_a = \eta (mc)^2 \, .
\ee
We shall employ the usual relativistic abbreviations,
\be
\bm{\beta}_{\bm{v}} \equiv \frac{\bm{v}}{c} \qqd \gamma_{\bm{v}} \equiv \frac{1}{\sqrt{1 - \bm{\beta}_{\bm{v}}^2}} \ .
\ee
Also, we use the abstract index notation (see \cite{Wald}) in which the Latin indices denote a tensor, while the Greek indices denote the components of a tensor in a specific coordinate system (3-vectors are denoted by a bold symbol, or by Latin indices from the middle of the alphabet). This notation is used when one wants to emphasize whether some equation is a tensorial equality (one which is valid in all coordinate systems), or merely an equality valid in some \emph{particular} coordinate system and \emph{not necessarily} in others. For example, equation $\tensor{S}{^a_b} = \tensor{T}{^a_b}$ immediately implies that $\tensor{S}{^\mu_\nu} = \tensor{T}{^\mu_\nu}$ is valid in any coordinate system $\{x^\mu\}$, but the converse does not necessarily hold. Furthermore, basis vectors adapted to some coordinate system $\{x^\mu\}$ are denoted by $e^{a}_{(\mu)}$, where the parenthesis around the index ``$\mu$'' is used to remind us that this is a collection of vectors, not components of a single vector. 
 
\vspace{20pt}

\section{Kinematical examples}

\subsection{Energy, mass and all that}

One of the first lessons for students to learn about relativity is that the energy $E$ of particles is conserved but is not an invariant quantity, whereas the (rest) mass $m$ is not conserved but is an invariant quantity (see e.g.~\cite{GriffPart}, p.~97). The first part of this ``mantra'' can easily lead to a wrong conclusion that it is \emph{impossible} to write energy using some manifestly invariant expression. It can be seen in the following way that this is not the case. Suppose some observer moving with 4-velocity $\fv{o}^a$ is measuring the energy $E$ of some particle with 4-momentum $\fv{p}^a$. The outcome of such an experiment is given by
\be\label{eq:Epo}
E(o) = \eta\,\fv{p}^a \fv{o}_a \, .
\ee
The fact that the energy is here formally written using manifestly invariant expression on the rhs should be taken with a pinch of salt. The ``catch'' lies in the fact that this scalar quantity is, obviously, \emph{observer dependent}, which is emphasized by writing the energy as a function of $\fv{o}^a$. Here lies the central point of the common assertion that ``the energy is not an invariant quantity'': two distinct observers, moving with 4-velocities $\fv{o}^a$ and $\fv{u}^a$ (such that $\fv{o}^a \ne \fv{u}^a$) will measure different energies of the same particle,
\be
E(o) = \eta\,\fv{p}^a \fv{o}_a \ne \eta\,\fv{p}^a \fv{u}_a = E(u) \, .
\ee
The energy of the particle in relativistic physics is usually described as the ``zeroth'' or the ``time'' component of the 4-momentum $\fv{p}^a$. Transformation of the energy then follows from the transformation properties of the 4-momentum. However, here one has to bring to mind that the physical meaning behind the coordinate or frame change is the change of the observer who performs the measurement of the energy (and other physical observables). Equation (\ref{eq:Epo}) makes this relation more tangible since the observer is explicitly present there. A naive reader might be tempted to call energy a ``scalar quantity'', however, this coordinate invariance is delusive: for example, switching to a boosted coordinate system without the corresponding change of the observer\footnote{Several remarks regarding this point can be found in \cite{Herr}.} would be a mathematically well-defined, but a physically meaningless procedure.

\bigskip

Furthermore, assuming that we are observing a free massive particle with the rest mass $m$ and 4-velocity $v^a$, its energy can be split into the rest energy and the kinetic energy,
\be
E(o) = mc^2 + T(o) \, ,
\ee 
where
\be
mc^2 = \eta\,\fv{p}^a \fv{v}_a \txand T(o) = \eta\,\fv{p}^a (\fv{o}_a - \fv{v}_a) \, .
\ee
It is clear that the kinetic energy $T(o)$ is again an observer dependent quantity, whereas the rest mass $m$ is manifestly \emph{invariant} and \emph{observer independent}. In other words, all observers should agree about the rest mass of a given particle. Note that one can also look upon the energy as being comprised of two contributions: the motion-induced kinetic energy and the rest of the energy, which does not depend on the motion of the particle and is conveniently called the rest energy. Dividing the rest energy by $c^2$, we obtain an observer independent Lorentz scalar $m$ without explicitly invoking the fact that it is related to the square of the 4-momentum. We shall later see that the similar reasoning also holds for the spin of the particle.

\bigskip

We shall proceed with a short overview of the most important kinematic equations in special theory of relativity, keeping the observer's 4-velocity $\fv{o}^a$ explicit. First we note that, given an observer in the Minkowski spacetime, there is a natural choice of a family of coordinate systems at each point of its trajectory, namely those in which the observer is at rest. This choice is not unique because we can always perform a spatial rotation of the coordinate system, thus keeping the observer at rest. In any such \emph{observer adapted} coordinate systems, the 4-velocity of the observer will have the following components
\be
\fv{o}^\mu = (c, \bm{0}) \, .
\ee
It is customary to choose one of the observer adapted coordinate systems and call it laboratory (LAB) frame. Suppose now that we have two particles moving through the Minkowski spacetime with 4-velocities $\fv{u}^a$ and $\fv{v}^a$. We want to find their relative speed $w$, that is, the modulus of the velocity of the other particle measured by either of them. The standard relativistic formula (see e.g.~\cite{LL2}, section 12) reads
\be\label{eq:betaw}
\bet_{\bm{w}}^2 = \frac{(\bet_{\bm{u}} - \bet_{\bm{v}})^2 - (\bet_{\bm{u}}\times\bet_{\bm{v}})^2}{(1 - \bet_{\bm{u}}\cdot\bet_{\bm{v}})^2} \, ,
\ee
and it is straightforward but tedious to check that the rhs of this equation is indeed (Lorentz-) invariant. This computational inconvenience can be made more tractable using the covariant prescription. In the LAB frame the 4-velocities of these two particles are given by
\be
\fv{u}^\mu = (\gamma_{\bm{u}} c, \gamma_{\bm{u}} \bm{u}) \qqd \fv{v}^\mu = (\gamma_{\bm{v}} c, \gamma_{\bm{v}} \bm{v}) \, .
\ee
If we choose the coordinate system in which one of the particles, say ``$\fv{u}$'' is at rest, we have
\be
\fv{u}^{\mu'} = (c,\bm{0}) \qqd \fv{v}^{\mu'} = (\gamma_{\bm{w}} c,\gamma_{\bm{w}} \bm{w}) \, .
\ee
Taking the scalar product between 4-vectors $\fv{u}^a$ and $\fv{v}^a$ one gets 
\be\label{eq:gammaw}
\gamma_{\bm{w}} = \eta\,\frac{\fv{u}_a \fv{v}^a}{c^2} \, .
\ee
The equation (\ref{eq:gammaw}) is manifestly invariant, as well as symmetric with respect to the 4-velocities $\fv{u}^a$ and $\fv{v}^a$, and can be used to extract the velocity of the particle $\fv{v}$ measured by the observer $\fv{u}$ and vice versa. Namely, by evaluating the rhs of (\ref{eq:gammaw}) in the LAB frame we immediately have
\be
\gamma_{\bm{w}} = \gamma_{\bm{u}} \gamma_{\bm{v}} (1 - \bet_{\bm{u}}\cdot\bet_{\bm{v}}) \, ,
\ee
and it is straightforward to check that (\ref{eq:betaw}) follows from here.

\bigskip

\subsection{Doppler effect}

Another elusive example in relativistic kinematics is the Doppler effect. Let $\fv{o}^a$ be the 4-velocity of an observer and $\fv{e}^a$ the 4-velocity of an emitter of monochromatic photons. We want to find the relation between the frequency $\omega_e$ of the emitted photon, measured in the rest frame of the emitter, and the frequency $\omega_o$ of the same photon measured in the observer's rest frame. Unfortunately, textbooks often tend to rederive various formulae for different types of this phenomenon (e.g.~\cite{LL2,Morin,Jackson}), without providing the easy--to--remember ``master formula''. Such can be achieved easily through covariant transcription, as follows. The 4-momentum $p^a$ and 4-wave vector $k^a$ of the photon are generally given by 
\be
\fv{p}^\mu = (\hbar\omega/c,\hbar\bm{k}) \qqd \fv{k}^a \equiv \fv{p}^a/\hbar \, .
\ee
Since the photon is massless, its dispersion relation is given by
\be
\fv{p}^2 = 0 = \fv{k}^2 \, .
\ee
Note that $|\bm{k}| = \omega/c$. The frequency of the photon, measured by the observer and the emitter can be written succinctly as
\be
\omega_o \equiv \omega (\fv{o}) = \eta\,\fv{k}_a \fv{o}^a \qqd \omega_e \equiv \omega (\fv{e}) = \eta\,\fv{k}_a \fv{e}^a \, .
\ee
The Doppler redshift is usually expressed through an invariant, but observer dependent quantity $z$, defined via
\be\label{eq:Dopplerz}
z(o) + 1 \equiv \frac{\omega_e}{\omega_o} = \frac{\fv{k}_a \fv{e}^a}{\fv{k}_a \fv{o}^a} \, .
\ee
We shall now demonstrate how this elegant formula allows one to easily switch between different frames. If, measured from the LAB frame, the photon of frequency $\omega$ is emitted in the space direction defined by the unit 3-vector $\bm{\hat{s}}$ from the emitter moving with 3-velocity $\bm{v}$, we have 
\be
c\fv{k}^\mu = (\omega,\omega\bm{\hat{s}}) \qqd \fv{o}^\mu = (c,\bm{0}) \quad \textrm{and} \quad \fv{e}^\mu = (\gamma_{\bm{v}} c,\gamma_{\bm{v}}\bm{v}) \, .
\ee
Inserting these elements into equation (\ref{eq:Dopplerz}) one gets
\be\label{eq:Dopplerz2}
z + 1 = \gamma_{\bm{v}} \left( 1 - \bet_{\bm{v}}\cdot\bm{\hat{s}} \right) \, .
\ee
It is instructive to consider two special cases of the Doppler effect,

\begin{itemize}

\item Longitudinal Doppler effect: the photon is emitted along the direction of the emitter's velocity, $\bm{\hat{s}} = \mp \bm{\hat{v}}$, with the upper sign for emitter moving \emph{away} from the observer and the lower sign in the opposite case, so that (\ref{eq:Dopplerz2}) reads
\be
z + 1 = \gamma_{\bm{v}} \left( 1 \pm |\bet_{\bm{v}}| \right) = \sqrt{\frac{1 \pm |\bet_{\bm{v}}|}{1 \mp |\bet_{\bm{v}}|}} \ .
\ee

\bigskip

\item Transverse Doppler effect: the photon is emitted perpendicular to the direction of the emitter's velocity, $\bm{\hat{s}}\cdot\bm{\hat{v}} = 0$. This situation occurs if, for example, the emitter is moving along a circle at a constant distance from the observer at the center of the circle. Inserting this into (\ref{eq:Dopplerz2}) one immediately gets
\be
z + 1 = \gamma_{\bm{v}} \, .
\ee
Note that this formula cannot be applied in the case when the observer is moving along the circle around the emitter in the center. In the rest frame of the emitter,
\be
\fv{o}^{\mu'} = (\gamma_{\bm{o}} c, \gamma_{\bm{o}} \bm{o}) \qqd \fv{e}^{\mu'} = (c,\bm{0}) \, ,
\ee
we have $\bm{\hat{s}}\cdot\bm{\hat{o}} = 0$ at the moment of photon detection, so by making use of (\ref{eq:Dopplerz}) again,
\be
z + 1 = \frac{1}{\gamma_{\bm{o}}} \, .
\ee

\end{itemize}

\bigskip

\noindent
Further examples of uses of the equation (\ref{eq:Dopplerz}) for various gravitational redshifts can be found in \cite{Wald}. We now turn to some examples in classical electrodynamics. 

\vspace{20pt}

\section{Classical electrodynamics}

In the standard approach to electrodynamics, the electromagnetic field is described by the antisymmetric electromagnetic field tensor $F_{ab}$ (also known as the Faraday tensor). Evaluating this tensor in a specific inertial reference frame $\mathcal{R}$ in which $F_{ab} \to F_{\mu \nu}$ allows for an identification of the electric and magnetic 3-vector fields $\bm{E}$ and $\bm{B}$ \emph{measured in that frame} via the correspondence
\be
F_{\mu\nu} = \frac{\eta}{c} \left( \begin{array}{cccc} 0 & E_x & E_y & E_z \\ -E_x & 0 & -cB_z & cB_y \\ -E_y & cB_z & 0 & -cB_x \\ -E_z & -cB_y & cB_x & 0 \end{array} \right).
\ee
Since the reference frame $\mathcal{R}$ was arbitrary, the same relation must hold for any other inertial reference frame $\mathcal{R}'$ with $F_{\mu \nu} \to F_{\mu' \nu'}$, $\bm{E} \to \bm{E}'$ and $\bm{B} \to \bm{B}'$. As the two reference frames are related by a Lorentz transformation, it is straightforward to derive the Lorentz transformation of electromagnetic field via
\be\label{eq:LorentzE}
\bm{E}' = \gamma_{\bm{v}} \left(\bm{E} + \bet_{\bm{v}} \times c\bm{B}\right) - (\gamma_{\bm{v}} - 1) (\bm{\hat{v}} \cdot \bm{E}) \bm{\hat{v}} \, ,
\ee
\be
c\bm{B}' = \gamma_{\bm{v}} \left(c\bm{B} - \bet_{\bm{v}} \times \bm{E}\right) - (\gamma_{\bm{v}} - 1) (\bm{\hat{v}} \cdot c\bm{B}) \bm{\hat{v}} \, ,
\ee
where $\bm{v}$ is the velocity of the reference frame $\mathcal{R}$ with respect to the reference frame $\mathcal{R}'$. These transformation properties are in agreement with all known experiments (see e.g.~\cite{Jackson}).

As an alternative approach\footnote{Examples of its use can be found e.g.~in \cite{Wald,C89,Heusler,Smo}.}, one can explicitly introduce the observers which measure electric and magnetic fields in the following way: for an observer moving through spacetime with 4-velocity $o^a$, define the electric and magnetic 4-vectors as
\be
E^a(o) = F^{ab} o_b, \txand c B^a(o) = -\,\sF^{ab} o_b \, ,
\ee
with $\sF_{ab} = \frac{1}{2}\,\epsilon_{abcd} F^{cd}$ being the Hodge dual of the Faraday tensor and $\epsilon_{0123} = 1$. Since, due to contraction of symmetric and antisymmetric tensors, we have
\be
o_a E^a (o) = F^{ab} o_a o_b = 0 \ , \quad \textrm{and} \quad o_a B^a(o) = -\frac{1}{c}\,\sF^{ab} o_a o_b = 0 \ ,
\ee
the electric and magnetic 4-vectors have three independent components each. In the reference frame $\mathcal{R}$ where $o^{\mu} = (c,\bm{0})$, we have 
\be
E^{\mu}(o) = (0,\bm{E}) \txand B^{\mu}(o) = (0,\bm{B}) \, .
\ee
By definition, the electric and magnetic field 4-vectors are observer dependent and, therefore, \emph{adapted for use in a specific reference frame}. Nevertheless, they can be used in the construction of observer independent quantities. For example, the Faraday tensor can be expressed as
\be\label{eq:FEB}
F_{ab} = \frac{\eta}{c^2} \left(E_a (o) o_b - E_b (o) o_a - \epsilon_{abcd} o^c cB^d(o)\right) \, ,
\ee
for any timelike $o^a$ with $o_a o^a = \eta c^2$. It also holds that
\be
E_a(o) E^a(o) - cB_a(o) cB^a(o) = \frac{\eta c^2}{2} F_{ab}F^{ab} \, ,
\ee 
and
\be
E_a(o)cB^a(o) = -\frac{\eta c^2}{4}\,F_{ab} \sF^{ab} \, .
\ee
These relations generalize the well known expressions for the two Lorentz invariants of the electromagnetic field.

Moreover, the Lorentz force law for a charged particle $q$ moving in an elec\-tro\-mag\-ne\-tic field reads
\be
\frac{\d}{\d \tau}\,m u^a = q \tensor{F}{^a_b} u^b \equiv q E^a (u) \, ,
\ee
where $u_a$ is the particle's 4-velocity and $\tau$ is its proper time. This expression is explicitly observer independent since the only 4-velocity present is that of the particle on which the force acts, not of the observer who is merely a passive spectator. Furthermore, depending on a reference frame in which one writes the equation, $E_a (u)$ stands for the combination of both electric and magnetic 3-vectors (the familiar 3-vector representation of Lorentz law). Only in the particle's rest frame $u^{\mu} = (c,\bm{0})$ do we have a purely electric field $E^{\mu}(u) = (0, \bm{E})$. What this means is that all charged particles (without spin) can experience for themselves only the effect of an electric field. In other words, classical charged particles have no direct notion of the existence of magnetic fields (this fact was stressed in \cite{Crater,VMC}).

To conclude this section, it is worth mentioning that the whole business of introducing the electric and magnetic 4-vectors is quite subtle. Given the Faraday tensor $F_{ab}$ and two observers $o$ and $o'$, one can construct two different electric 4-vectors $E^a(o) = F^{ab}\,o_b$ and $E^a(o') = F^{ab}\,o'_b$ (similarly for the magnetic 4-vectors) so that $E^{\mu}(o) = (0, \bm{E})$ and $E^{\mu'}(o') = (0, \bm{E}')$. The 4-vector $E^a(o)$ is related to the electric field 3-vector as measured by $o$, and the same holds for $E^a(o')$ and the observer $o'$. Since the electric field $\bm{E}$ alone does not transform under the irreducible representation of a Lorentz group (it does so in the combination with the magnetic field $\bm{B}$ !) we must not expect the relation of the type
\be\label{eq:wrongE}
E^{\mu'}(o') = \tensor{\Lambda}{^{\mu'}_{\nu}} E^{\nu}(o)
\ee
to hold. Explicitly, the only Lorentz transformation that satisfies
\be
\left(\begin{array}{c} 0 \\ E'^{\,i'} \end{array}\right) = \left( \begin{array}{cc} \tensor{\Lambda}{^{0'}_0} & \tensor{\Lambda}{^{0'}_i} \\  \tensor{\Lambda}{^{i'}_0} & \tensor{\Lambda}{^{i'}_i} \end{array} \right) \left(\begin{array}{c} 0 \\ E^i \end{array} \right),
\ee
is the 3-rotation transformation $\tensor{\Lambda}{^{\mu'}_\nu}=\tensor{R}{^{\mu'}_\nu}$.

Furthermore, if the relation (\ref{eq:wrongE}) were to hold for Lorentz boosts, then the electric and magnetic field would be ``uncoupled'' with respect to the Lorentz transformation and this would imply, amongst other things, that moving electrons produce no magnetic field. The transformation (\ref{eq:wrongE}) is mathematically well-defined, but physically meaningless, being at odds with the original motivation for introducing the 4-vector $E^a(o)$. Instead, the following \emph{does} hold
\be\label{eq:goodE}
E^{\mu'}(o) = \tensor{\Lambda}{^{\mu'}_\nu} E^{\nu}(o) \, ,
\ee
due to the fact that $E^a(o)$ is a genuine 4-vector. The relation (\ref{eq:goodE}), however, still does not tell us anything about the measurement of the observer $o'$. This information can be obtained via relation (\ref{eq:FEB}), by contraction with 4-velocity $o'^a$ and evaluation in the ``primed'' coordinate system 
\be
\begin{array}{l}
E^{\mu'}(o') = \tensor{F}{^{\mu'}_{\nu'}} o'^{\,\nu'} \\
\qquad \quad \, = \frac{\eta}{c^2} \left( E^{\mu'}(o) o_{\nu'} - E_{\nu'}(o) o^{\mu'} - \tensor{\epsilon}{^{\mu'}_{\nu'}_{\rho'}_{\sigma'}} o^{\rho'} cB^{\sigma'}(o) \right) o'^{\,\nu'},
\end{array}
\ee
which, together with (\ref{eq:goodE}), reduces to (\ref{eq:LorentzE}).

The potential for confusion in the above reasoning is best confirmed by the series of published papers (see \cite{Ivezic} and references therein) errorneously claiming the relation (\ref{eq:wrongE}) to be the true Lorentz transformation of electric field. While this is cleary nonsensical, the authors of the cited papers insist that the physical (measurable) quantities must necessarily transform under the irreducible representation of a Lorentz group, which is known not to be the case (see e.g.~\cite{Jackson} and references therein).  

\vspace{20pt}

\section{Thomas precession}

Our final example is also the most elaborate, the well-known Thomas precession. The textbook derivations of this effect \cite{Jackson,Gold,Padma} are often incomplete, lack precision and leave much to be desired. A somewhat detailed analysis can be found in \cite{Garg}, with a particular emphasis on the role of the observers\footnote{A deeper, geometric picture behind the Thomas precession and its ``relative'', Foucault precession, is described in \cite{Krivo}.}. In the standard presentation, the spin of the electron orbiting around the nucleus is found to precess due to a combination of the relativistic and the Coriolis effects. To derive the result, one must use several different reference frames, the distinction of which is often blurred in the nonrelativistic approximation, so that in the end it is unclear in which of the reference frames would the effect be measurable. Our goal is to rederive the Thomas precession, taking into account the observers so that it is clear at each moment which observer measures each effect. However, before we begin the derivation, we shall briefly discuss the observer (in)dependence of angular momentum.

The angular momentum of a particle is given by the antisymmetric angular momentum tensor $J_{ab}$. In a specific reference frame $\mathcal{R}$, we have $J_{ab} \to J_{\mu \nu}$ with
\be
J_{\mu\nu} = \left( \begin{array}{cccc} 0 & -K_x & -K_y & -K_z \\ K_x & 0 & J_z & -J_y \\ K_y & -J_z & 0 & J_x \\ K_z & J_y & -J_x & 0 \end{array} \right).
\ee
Here, $\bm{K}$ is the boost 3-vector describing the movement of the particle's center of mass, while $\bm{J}$ is the angular momentum 3-vector (see e.g.~\cite{LL2}, pp.~44--45). Both of these vectors depend not only on the chosen reference frame in which they are evaluated (measured) but also on the choice of the origin of coordinate system.

Similar to the definition of the magnetic 4-vector, the angular momentum 4-vector measured by an observer $o$ is
\be
J^a (o) = \frac{1}{c}\,\sJ^{a b}o_b \, .
\ee
We can now split the total angular momentum into the orbital (that is, motion-induced) angular momentum,
\be
L^a (o) = \frac{1}{c}\,\sJ^{a b}(o_b - u_b) \, ,
\ee
and the spin (intrinsic angular momentum),
\be
S^a = \frac{1}{c}\,\sJ^{a b}u_b \, ,
\ee
where $u^a$ is the 4-velocity of the particle and immediately we have $u^a S_a = 0$. The spin 4-vector is obviously origin independent. It is easily seen that the separation of total angular momentum into orbital and spin angular momentum 
\be
J^a(o) = L^a(o) + S^a
\ee
is observer dependent.

From the above analysis, it is clear that the spin 4-vector $S^a$ is observer independent. However, what is usually meant by the spin of the particle is the spin 3-vector $\bm{s}$ which has to be related in some way to the 4-vector $S^a$. It is defined as the spatial part of $S^a$ as measured by the observer in the particle's rest frame $\mathcal{K'}$. Therefore,
\be
S^a \to S^{\mu} \equiv (0, \bm{s}) \, .
\ee
With this definition, the standard notion of a spin 3-vector is observer dependent since the components of $S^a$ in any other reference frame become
\be
S^a \to S^{\mu'} = (S^0, \bm{S}) \, ,
\ee
with $\bm{S}$ being the spin of the particle as measured by an observer moving with relative velocity $\bm{v}$ and $S^0 = \bet_{\bm{v}} \cdot \bm{S}$ being the corresponding helicity of the particle. The observer dependence of the 3-vector spin comes from the fact that under the Lorentz transformation we have (see \cite{Jackson}, section 11.11)
\be\label{eq:sSrel}
\bm{S} = \bm{s} + \frac{\gamma_{\bm{v}}^2}{\gamma_{\bm{v}}+1}\,(\bet_{\bm{v}}\cdot \bm{s}) \bet_{\bm{v}} \, ,
\ee
i.e.~the 3-spin undergoes an inverse Lorentz contraction (a dilation). One may ask is it meaningful to speak of a spin and not of the total angular momentum of a particle when one observes it in an arbitrary reference frame? The answer is --- yes. This is intuitively clear, since we can always differentiate between the rotation about its own axis and orbital motion. The point here is that the change of the observer changes both the orbital angular momentum (this is also true in nonrelativistic mechanics) and the spin (a relativistic effect) of the particle.

Let us now return to the problem of Thomas precession. We are interested in the properties of a \emph{classical} electron revolving around the nucleus (that is, we describe the electron as a relativistic point-like particle). Let us introduce the LAB reference frame $\mathcal{K}$ as an inertial reference frame in which the nucleus is at rest. This is the reference frame in which we perform the experiment and observe all relevant effects. The comoving (CM) frame $\mathcal{K}'$, which we take to be the rest frame of the electron, is equally important. Since the electron is in noninertial motion, this reference frame is also noninertial. To this end, we identify the CM frame with the set of inertial frames momentarily comoving with the electron. This effectively means that the LAB and CM frames are related by a (proper) time dependent Lorentz transformation. The most general Lorentz transformation can be uniquely separated into a pure Lorentz boost followed by a 3-rotation. The separation introduces another, noninertial (BOOSTed) reference frame $\widetilde{\mathcal{K}}$ so that the following holds
\be
\mathcal{K} \ {\buildrel \textrm{\tiny{BOOST}} \over \longrightarrow} \ \widetilde{\mathcal{K}} \ {\buildrel \textrm{\tiny{ROT}} \ \over \longrightarrow} \ \mathcal{K}' \, .
\ee
Note that this is the first time we are considering two reference frames (observers) that share the same 4-velocity as being different, since their spatial axes differ by a 3-rotation. In general, an observer is uniquely determined by all of the basis vectors of her/his reference frame, and not only her/his 4-velocity. In what follows, $o^a$ denotes the 4-velocity of the LAB observer, $u^a$ denotes the electron's 4-velocity and $\omega^a$ the angular velocity of the CM frame with respect to the BOOST frame, so that $\omega^a \to \omega^{\mu'} = (0,\bm{\omega})$. Also, $a^b = \dot{u}^b$ is the electron's 4-acceleration, where the dot represents the derivative with respect to $\tau$, the electron's proper time.

Care must be taken when considering noninertial frames $\mathcal{K}'$ and $\widetilde{\mathcal{K}}$ as the orthonormal basis vectors of these frames $e^a_{(\mu')}$ and $e^a_{(\tilde{\mu})}$ are not constant during the motion and their evolution generates Coriolis-like terms in the equations of motion for physical quantities (see e.g.~\cite{Gold}). The reason for this peculiar effect is that by definition, a given observer will see her/his basis vectors as fixed. Therefore, the rate of change of the spin, as well as any other, 4-vector $S^a = S^{\mu}e^a_{(\mu)} = S^{\mu'}e^a_{(\mu')}$ is perceived differently by the LAB and CM observer. We have, respectively,
\be
\dot{S}^a(o) = \frac{\d S^{\mu}}{\d \tau}\,e^a_{(\mu)} \quad \textrm{and} \quad \, \dot{S}^a(u) = \frac{\d S^{\mu'}}{\d \tau}\,e^a_{(\mu')} \ ,
\ee
so that
\be
\dot{S}^a(o) \equiv \dot{S}^a - S^\mu \dot{e}^a_{(\mu)} \quad \textrm{and} \quad \, \dot{S}^a(u) \equiv \dot{S}^a - S^{\mu'} \dot{e}^a_{(\mu')} \ .
\ee
By now it should be clear that $\dot{S}^a(o)$ and $\dot{S}^a(u)$ represent two different 4-vectors, not the same 4-vector represented in two different reference frames.

It can be shown that the most general relation between the time evolution of the spin 4-vector in the LAB and CM frames is given by\footnote{The straightforward derivation of this formula is quite lengthy but elementary. We give an outline of the derivation in the appendix.}
\be
\label{relacija}
\dot{S}^a(o) = \dot{S}^a(u) + \eta\,\tensor{\Omega}{^a_b} S^b \, .
\ee
Here we have introduced an observer dependent (for notational simplicity, we keep the observer dependence implicit) Coriolis-like tensor 
\be
\label{omega}
\Omega^{ab} = \Omega_{\mathrm{FW}}^{ab} + \Omega_{\mathrm{T}}^{ab} + \Omega_{\mathrm{R}}^{ab}
\ee
which consists of three parts:

\begin{itemize}
\item the Fermi-Walker term,
\be
\Omega_{\mathrm{FW}}^{ab} = \frac{1}{c^2} \left(a^a u^b - a^b u^a\right) \ ,
\ee

\bigskip

\item the Thomas term,
\be
\Omega_{\mathrm{T}}^{ab} = \frac{1}{c^2}\,\frac{a^a o_\perp^b - a^b o_\perp^a}{1 + \frac{\eta}{c^2}\,u^c o_c} \qqd o_\perp^a = o^a - \eta\,\frac{u^c o_c}{c^2}\,u^a \ ,
\label{Thmeq}
\ee

\bigskip

\item the rotation term,
\be
\Omega_{\mathrm{R}}^{ab} = \frac{1}{c}\,\epsilon^{abcd} u_c \omega_d \ .
\label{Roteq}
\ee
\end{itemize}

\noindent
The latter two of these three terms are observer dependent due to the presence of the observer's 4-velocity $o^a$ and the arbitrary angular velocity 4-vector $\omega^a$, which is an indirect consequence of introducing several reference frames.

The arbitrary nature of $\omega^a$ comes from the fact that, so far, we have just been doing mathematics. Equation (\ref{relacija}) is a mathematical statement relating the evolution of vectors in different reference frames. Therefore, the choice of $\omega^a$ fixes the relation between those reference frames and vice versa. In the case at hand, we fix $\omega^a$ as follows. Our starting reference frame (LAB) was inertial so that we have
\be
\dot{e}^a_{(\mu)} = 0 \ \Rightarrow \ \dot{S}^a(o) \equiv \dot{S}^a\,.
\ee
Similarly, the CM frame was chosen as the electron rest frame so that in the absence of external torques we have
\be
\dot{S}^a(u) \equiv 0\, .
\ee
Finally, we employ the principle of relativity which says that all inertial observers must agree on the evolution of the electron, i.e. the evolution must be observer independent. In other words, we demand
\be
\Omega_{\mathrm{R}}^{ab} + \Omega_{\mathrm{T}}^{ab} = 0 \, ,
\ee
which, expressing Eqs. (\ref{Thmeq}) and (\ref{Roteq}) in the LAB frame, gives the angular velocity of CM frame with respect to the BOOST frame
\be
\omega^{\mu} = \frac{1}{c^2}\,\frac{\gamma_{\bm{v}}^3}{1+\gamma_{\bm{v}}}\,(0, \bm{a} \times \bm{v}) \, .
\ee
This is the famous Thomas precession, where $\bm{v}$ and $\bm{a}$ are the velocity and the acceleration of the electron in the LAB frame.

We have learned two things. The first is that the rest frame of the electron in arbitrary motion is not simply the boosted laboratory (inertial) frame but that one must also rotate the boosted axes in accord to Thomas prescription. The second point is that the evolution of a torque-free spin is the observer independent Fermi-Walker transport (see e.g.~\cite{MTW}, pp.~170--172)
\be
\frac{\d S^a}{\d \tau}= \frac{\eta}{c^2} \left(a^a u_b - a_b u^a\right) S^b \, .
\ee
This is the equation one has to solve for the motion of the electron around the nucleus. To find out how different observers see the electron, we merely have to evaluate the above equation in an appropriate reference frame. From this point on, we shall use the coordinate time of a particular reference frame as a parameter of the evolution. For the CM frame, we have (with $\bm{s} \equiv \bm{S'}$)
\be
\frac{\d \bm{s}}{\d t} = \bm{0} \, ,
\ee
as expected by the definition of the rest frame. For the BOOST frame, we have
\be
\frac{\d \bm{\tilde{S}}}{\d t} = \bm{\omega} \times \bm{\tilde{S}} \, ,
\ee
an equation typically derived as the Thomas precession equation. However, this equation is nothing more than the Coriolis theorem relating the two frames differing by a rotation and is purely nonrelativistic in origin. Also, by assumption, there is no measuring apparatus in the BOOST frame so that no physical observer can observe this motion. Finally, in the LAB frame, the spin Fermi-Walker transport reads
\be
\frac{\d \bm{S}}{\d t} = \frac{\gamma_{\bm{v}}^2}{c^2}\left(\bm{a} \cdot \bm{S}\right)\bm{v} \, ,
\ee
where the relation (\ref{eq:sSrel}) holds. It is seen that in the LAB frame, the motion of the electron is far more complicated than a mere precession. For an electron in a circular motion in the $xy$ plane with angular velocity $\omega$ and the initial condition $\bm{S}(t=0)=(S_{0x},S_{0y}, S_{0z})$ it is found that (see \cite{LPPT}, problem 11.7)
\bea
S_x(t) &= S_{0x} \cos (\gamma_{\bm{v}} - 1) \omega t + S_{0y} \sin (\gamma_{\bm{v}} - 1) \omega t \nonumber \0\\
&+(\gamma_{\bm{v}} - 1) (S_{0x} \sin \omega t - \frac{1}{\gamma_{\bm{v}}} S_{0y} \cos \omega t) \sin \gamma_{\bm{v}} \omega t \, , \\ 
S_y(t) &= - S_{0x} \sin (\gamma_{\bm{v}} - 1) \omega t + S_{0y} \cos (\gamma_{\bm{v}} - 1) \omega t \nonumber \0\\
&-(\gamma_{\bm{v}} - 1) (S_{0x} \cos \omega t + \frac{1}{\gamma_{\bm{v}}} S_{0y} \sin \omega t) \sin \gamma_{\bm{v}} \omega t \, , \\
S_z(t) &= S_{0z} \, .
\eea
In the LAB frame, not only does the electron spin change its direction, but it also changes its magnitude. Why is this fact rarely mentioned in textbooks? It is because the average velocity of the electron in a hydrogen atom is nonrelativistic, so that the LAB frame is practically indistinguishable from the BOOST frame and the leading order relativistic effect is indeed the Thomas precession with angular velocity
\be
\omega_{\mathrm{T}} = (\gamma_{\bm{v}} - 1)\omega \approx \frac{1}{2c^2}\left| \bm{a} \times \bm{v} \right| \, .
\ee

\vspace{20pt}

\section{Final remarks}

Although it is always possible to conceive some practical, reduced notation, filled with hidden information, this choice will usually result in loss of pedagogical clarity. Such an example is the widespread custom to define various tensors and use different coordinate systems without the explicit reference to the pertaining observers. However, as can be seen from all the examples presented throughout this paper, it is impossible to stress the role of the observer in various aspects of invariance enough.

\vspace{20pt}

\appendix

\section{Derivation of equations (\ref{relacija}) and (\ref{omega})}

Let $o$ (denoted by Greek indices) and $u$ (denoted by primed Greek indices) be the two observers measuring the evolution of some 4-vector $S^a = S^{\mu} e^a_{(\mu)} = S^{\mu'} e^a_{(\mu')}$. The most general relation between the observers is a Lorentz transformation $\Lambda(\tau)$ that depends on some evolution parameter $\tau$ which will be implicitly understood in the following. The components of $S^a$ in the two reference frames are related by
\be
S^{\mu'} = \tensor{\Lambda}{^{\mu'}_{\mu}} S^{\mu} \qqd S^{\mu} = \tensor{\Lambda}{^{\mu}_{\mu'}} S^{\mu'}\, ,
\ee
and the similar relation also holds for the basis vectors,
\be
e^a_{(\mu')} = \tensor{\Lambda}{^{\mu}_{\mu'}} e^a_{(\mu)} \qqd e^a_{(\mu)} = \tensor{\Lambda}{^{\mu'}_{\mu}} e^a_{(\mu')}\, .
\ee
The evolution of 4-vector $S^a$ is observer dependent and reads
\be
\dot{S}^a(o) \equiv \dot{S}^a - S^{\mu} \dot{e}^a_{(\mu)} \qqd \dot{S}^a(u) \equiv \dot{S}^a - S^{\mu'} \dot{e}^a_{(\mu')}
\ee
for the $o$ and $u$ observer, respectively. Eliminating the $\dot{S}^a$ term in the above equations we obtain
\be
\dot{S}^a(o) = \dot{S}^a(u) + S^{\mu'} \dot{e}^a_{(\mu')} - S^{\mu} \dot{e}^a_{(\mu)}\, .
\ee
Using the Lorentz transformations, we can write the second term on the rhs as
\be
S^{\mu'} \dot{e}^a_{(\mu')} = S^{\mu'} \tensor{\dot{\Lambda}}{^{\mu}_{\mu'}} e^a_{(\mu)} + S^{\mu'} \tensor{\Lambda}{^{\mu}_{\mu'}} \dot{e}^a_{(\mu)} = \tensor{\dot{\Lambda}}{^{\mu}_{\mu'}} \tensor{\Lambda}{^{\mu'}_{\nu}} S^{\nu} e^a_{(\mu)} + S^{\mu} \dot{e}^a_{(\mu)}\, ,
\ee
i.e.
\be
\dot{S}^a(o) = \dot{S}^a(u) + \eta \tensor{\Omega}{^a _b} S^b\, ,
\ee
with
\be
\label{Lorentz}
\eta \tensor{\Omega}{^\mu _\nu} = \tensor{\dot{\Lambda}}{^{\mu}_{\mu'}} \tensor{\Lambda}{^{\mu'}_{\nu}}\, .
\ee

To explicitly determine the tensor $\tensor{\Omega}{^a _b}$, we use the fact that an arbitrary Lorentz transformation $\tensor{\Lambda}{^{\mu'}_{\mu}}$ can be expressed as a combination of a pure boost $\tensor{B}{^{\tilde{\mu}}_{\mu}}$ followed by a 3-rotation $\tensor{R}{^{\mu'}_{\tilde{\mu}}}$,
\be
\tensor{\Lambda}{^{\mu'}_{\mu}} = \tensor{R}{^{\mu'}_{\tilde{\mu}}} \tensor{B}{^{\tilde{\mu}}_{\mu}}\, .
\ee
Here we have introduced the intermediate reference frame with the coordinates $\{x^{\tilde{\mu}}\}$ which represents the boosted frame $o$ and shares the 4-velocity with the $u$ observer. The boost transformation $\tensor{B}{^{\tilde{\mu}}_{\mu}}$ is determined by the demand that the (normalized) 4-velocity of the $u$ observer is its zeroth basis vector
\be
e^a_{(0')} = \frac{1}{c}\, u^{\mu} e^a_{(\mu)} \ \Rightarrow \ \tensor{\Lambda}{^{\mu}_{0'}} = \frac{1}{c}\, u^{\mu} \ \Rightarrow \ \tensor{\Lambda}{^{0'}_{\mu}} = \frac{\eta}{c}\, u_{\mu}\,.
\ee
This implies
\be
\label{boost}
\tensor{B}{^{\tilde{\mu}} _\mu} = \left(
\begin{array}{cc}
\gamma & -\frac{\gamma}{c} v_j \\ -\frac{\gamma}{c}v_i & \delta_{ij} + \frac{\gamma-1}{\bm{v}^2} v_i v_j
\end{array}
\right)\, ,
\ee
where $\bm{v}$ is the 3-velocity of $u$ as measured by the observer $o$. For notational simplicity, we have put $\gamma \equiv \gamma_{\bm{v}}$. The 3-rotation transformation is trivial
\be
\label{rot}
\tensor{R}{^{\mu '} _{\tilde{\mu}}} = \left(
\begin{array}{cc}
1 & 0_{\tilde{\jmath}} \\ 0_{\tilde{\imath}} & R_{i'\tilde{\jmath}}
\end{array}
\right)\, ,
\ee
where $R_{i'\tilde{\jmath}}$ is an arbitrary orthogonal matrix. Plugging equations (\ref{boost}) and (\ref{rot}) into the equation (\ref{Lorentz}), defining $R_{i' \tilde{k}} \dot{R}_{j' \tilde{k}} = - \epsilon_{i'j'k'} \omega_{k'}$ and performing some lengthy and tedious algebra, one finally obtains (\ref{omega}).

\ack

IS would like to acknowledge the financial support of the Croatian Ministry of Science, Education and Sport under the contract no.~119-0982930-1016. BK would like to acknowledge the financial support of the Croatian Ministry of Science, Education and Sport under the contract no.~098-0982390-2864. We would like to thank Sanjin Beni\'c, Tajron Juri\'c, Hrvoje Nikoli\'c and Branimir Radov\v ci\'c for their careful reading of the manuscript and many useful comments.

\vspace{25pt}

\section*{References}

\nocite{*} 
\bibliographystyle{unsrt}
\bibliography{incovobs}

\end{document}